\documentstyle[multicol,graphics,aps]{revtex}       

\title{The Thermopower of Quantum Chaos}
\author{S.F.\ Godijn, S.\ M\"{o}ller, H.\ Buhmann, and L.W.\ Molenkamp}
\address{2. Physikalisches Institut, RWTH-Aachen, Templergraben 55, D-52056 Aachen, Germany}

\author{S.A. van Langen\cite{byline}}
\address{Instituut-Lorentz, Leiden University, P.O. Box 9506, 2300 RA Leiden, 
The Netherlands\\
{(\rm Received 10 November 1998)}}

\begin{document}

\draft
\preprint{}

\maketitle
\begin{abstract}

The thermovoltage of a chaotic quantum dot is measured using a current heating technique. The fluctuations in 
the thermopower as a function of magnetic field and dot shape display a non-Gaussian distribution, in 
agreement with simulations using Random Matrix Theory. We observe no contributions from weak localization or 
short trajectories in the thermopower.

\end{abstract} 

\begin{multicols}{2}
 
The electrical conductance of small - characteristic size much smaller than the electron mean free path - 
confined electron systems (usually denoted as quantum dots) shows distinct fluctuations. These fluctuations 
display correlations as a function of an external parameter such as shape or magnetic field, which can be 
described in a statistical manner. The electrons can, in fact, be viewed as billiard balls moving in a 
classically chaotic system where many random reflections at the system walls occur. Because of the wave-like 
nature of the electrons, quantum mechanics is needed to describe these systems fully. Chaos in quantum dots 
has been investigated \cite{Beenakker,Marcus,Chan} in conductance measurements but the analysis turns out to 
be difficult. So-called short trajectories \cite{Baranger96} and weak localization effects 
\cite{Beenakker,PluharBaranger93} add up to the signature of chaotic motion. Moreover, current heating of the 
electrons in the dot appears to be unavoidable in conductance measurements. Electron heating effects in the 
dot smear out the underlying chaotic statistics and therefore the observed fluctuations exhibit mostly a 
Gaussian distribution, although theory predicts non-Gaussian distributions when a small number of electron 
modes is admitted to the dot \cite{Efetov95}. Only when dephasing (modelled as extra modes coupling the dot 
to the environment) is included, Random Matrix Theory (RMT) \cite{Beenakker,BBB} gives a Gaussian 
distribution. Very recently, Huibers et al. \cite{Huibers} observed small deviations from a Gaussian 
distribution in conductance measurements. However, other transport properties calculated from these data 
exhibit again Gaussian distributions in contrast to theoretical predictions.

An alternative for the conductance measurements pursued so far (which inherently are accompanied by electron 
heating inside the dot) is to investigate the thermoelectric properties of a system. Thermopower measurements 
have already been used to study semiconductor nanostructures like quantum 
point-contacts\cite{Molenkamp90} and quantum dots in the Coulomb blockade regime\cite{Molenkamp94,Moeller}. 
The thermopower $S$ measures directly the parametric derivative of the conductance, $S \propto G^{-1} 
\partial G / \partial X$ with $X=E$ (energy), and thus yields both similar and additional 
information on the electron transport processes as can be obtained from conductance measurements. The 
distribution of parametric derivatives ($X= E, B, {\rm shape},\ldots$) of the conductance of a quantum dot is 
the subject of recent RMT-investigations \cite{Fyodorov,Brouwer97b}. The probability distribution for the 
thermopower is again expected to be non-Gaussian for chaotic conductors, exhibiting cusps at zero amplitude 
and non-exponential tails \cite{Brouwer97b,Vanlangen}.

In this paper, we present magneto-thermopower measurements of a statistical ensemble of chaotic quantum dots. 
The observed thermopower fluctuations show a non-Gaussian distribution. We present a numerical fit based on 
RMT which describes the experimental data. We demonstrate that effects like short trajectories, weak 
localization and dephasing are absent in thermopower measurements. 

\begin{figure}[hb]
\begin{center}
\resizebox{5cm}{7cm}{\includegraphics{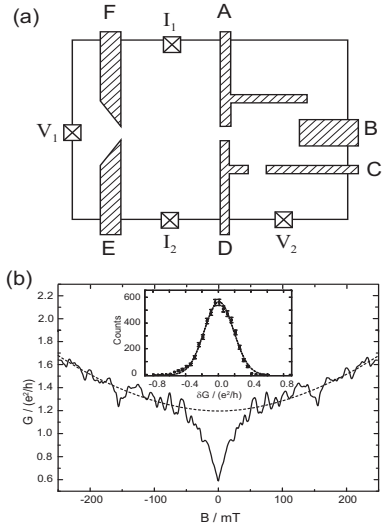}}
\end{center}
\begin{center}
\begin{minipage}{8cm}
\caption{(a) Schematic topview of the measured sample. The crosses denote the Ohmic contacts to the 2DEG; the 
hatched areas denote the gates. The heating current is applied between I1 and I2. The thermovoltage is 
measured between V1 and V2. The quantum dot is defined by applying a negative voltage to gates A,B,C and D. 
(b) Magnetoconductance of the dot, averaged over a large number of different dot configurations. The 
influence of short trajectories is characterized by the dashed line. Inset: Conductance distribution for 
$|B|\geq 50$ mT.}\label{sample}
\end{minipage}
\end{center}
\end{figure}

In Fig.~\ref{sample}a the measured device is shown schematically. A quantum dot (lithographic size $800$ nm 
$\times$ $700$ nm) is electrostatically defined (gates A, B, C and D) in a standard, high-mobility 
2-dimensional electron-gas (2DEG) in a GaAs-(Al,Ga)As heterostructure. The 2DEG has a mobility of $\mu 
\approx 10^6$ cm$^2$ (Vs)$^{-1}$ for an electron density of $3.4 \times 10^{11}$ cm$^{-2}$ at $4.2$~K. A 
2~$\mu$m wide and a 20 $\mu$m long electron-heating channel is defined next to the quantum dot (gates A, D, E 
and F). The sample is kept at 40 mK in a dilution refrigerator equipped with a superconducting magnet. 
Transport measurements are performed using standard phase-sensitive techniques. For reasons of comparison all 
data shown in this paper were obtained from the same sample. We have obtained similar results in several  
other devices.


Conductance data are shown in Fig.~\ref{sample}b. The graph is the magnetoresistance of the dot, averaged 
over a large number of different configurations\cite{average}. Both point contacts leading to the dot were 
adjusted to a conductance $G=4e^2/h$ corresponding to two spin-degenerate modes in the point contacts. As in 
Ref.~\cite{Chan}, an ensemble of configurations was created by repeatedly changing the voltage on gate B by a 
small amount ($\delta V_g^B=10$ mV). As is evident from the figure, apart from chaotic conductance 
fluctuations also the signatures of weak localization (sharp peak around $B=0$ T) and short trajectories 
(characterized by conductance background that exhibits a polynominal dependence on magnetic 
field\cite{Marcus,Chan}, Fig.~1b, dashed line) are clearly observed. In order to extract the statistics of 
the fluctuations, the conductance measurements were corrected for these features. The resulting distribution 
of the conductance fluctuations is shown in the inset of Fig.~\ref{sample}b (for magnetic fields larger than 
50 mT). A Gaussian function fits these data well. 

 
By passing a low-frequency (13 Hz) current through the ohmic contacts I$_1$ and I$_2$, the electron gas in 
the channel is heated ($T_e\propto I^2$) while the wide 2DEG regions remain in equilibrium with the lattice.  
The ac-heating current $I\approx 0.4$ $\mu$A is small enough to avoid lattice heating and we have ensured 
that we are in the regime of  linear response. The temperature difference between heating 
channel and electron reservoir induces a thermovoltage $V_{\rm th}$ across the chaotic dot. $V_{\rm th}$ can 
be measured, using quantum point-contact (QPC) E-F as a reference point, between ohmic contacts V$_1$ and 
V$_2$. $V_{\rm th}$ is then related to the thermopower of the dot, $S_{\rm dot}$, as

\begin{equation}\label{thermopower1}
V_{\rm th}=V_2-V_1=V_{\rm dot}-V_{\rm ref}=(S_{\rm dot}-S_{\rm ref})\Delta T
\end{equation}

Here, QPC E-F is adjusted such that its thermopower, $S_{\rm ref}$, is minimal and constant for all 
measurements. As in the conductance measurements, the transmittance of QPCs A-D and C-D was adjusted to 
$G=4e^2/h$. Again, varying the voltage applied to gate B, $V_g^B$, was used to change the shape of the dot. 

\begin{figure}[hb]
\begin{center}
\resizebox{7cm}{7cm}{\includegraphics{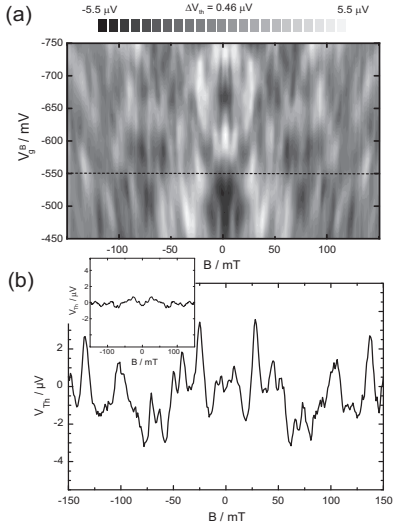}}
\end{center}
\begin{center}
\begin{minipage}{8cm}
\caption{(a) Grayscaleplot of the thermovoltage of the quantum dot as a function of magnetic field (X-Axis) 
and of gate voltage on gate B (Y-Axis). The gate voltage is changed by a constant small amount of $\delta 
V_g^B = 10$ mV for each magnetic field sweep. The light areas denote a large (maximum 5.5 $\mu V$) positive 
thermovoltage, the dark areas a large (maximum -5.5 $\mu V$) negative thermovoltage. (b) Individual 
thermovoltage trace for $V_g^B=-550$ mV [dashed line Fig.~(a)]; Inset: Thermovoltage, corrected for the 
temperature variation of the heating channel and averaged over a large number of dot configurations. 
Signatures of weak localization and short trajectories are absent.}\label{colorplot}
\end{minipage}
\end{center}
\end{figure}

In Fig.~\ref{colorplot}a a grayscaleplot is shown of the transverse voltage $V_{\rm th}$ as a function of 
magnetic field and $V_g^B$. The step in gate voltage between two successive magnetic-field sweeps was $\delta 
V_g^B = 10$ mV. The magnetic field range was limited to $|B| \leq$ 150 mT to avoid the regime where the 
quantum Hall effect becomes dominant. The characteristic fluctuations are stable in time and well
reproducible. As an example, the trace for $V_g^B=-550$ mV is plotted separately in Fig.~2b. The fluctuations 
are symmetric around $B=0$ T with a zero mean.


Because the resistance of the electron heating channel is magnetic-field dependent due to classical 
(breakdown of the entrance resistance) and quantum (weak localization and conductance fluctuations) transport 
effects, also the electron temperature in Eq.~1 is somewhat dependent on magnetic field, $T_e = T_e(B)$. For 
the given device structure, this dependence can easily be determined experimentally, using the quantized 
thermopower of a point contact \cite{Molenkamp90}. The thermovoltage of QPC A-D, adjusted for maximum 
thermopower, is measured as a function of magnetic field while keeping $V_g^B=V_g^C=0$ V, i.e.\ without 
defining the dot, yielding $V_{\rm th,channel}=(S_{\rm AD}-S_{\rm ref})\Delta T(B)$. The variation in 
channel-temperature, which turns out to be only a few percent, is effective for each individual measurement. 
Thus, by dividing the dot thermovoltage by the QPC thermovoltage, this effect can be eliminated. We have 
verified that this calibration does not influence the statistics of the thermopower fluctuations.


The inset of Fig.~\ref{colorplot}b shows the temperature-variation corrected thermovoltage, averaged over all 
configurations. We show this averaged trace to illustrate that in contrast to the averaged conductance 
(Fig.~1b) signatures of weak localization and short trajectories are absent in the averaged 
thermopower\cite{average}. For the weak localization correction, this is readily understood when one realizes 
that the correction to the conductance for zero dimensional system is energy independent and any signs of 
weak localization in the thermopower should derive from the energy dependence of the phase-coherence length 
$l_\phi$ which is presumably small. (However, see Ref.~\cite{kearney} for experiments on weak localization 
thermopower in the two-dimensional quantum-diffusive transport regime). The 
absence of a signature of short trajectory effects in the thermovoltage measurements implies also a very weak 
energy dependence for the conductance correction due to these processes. Heuristically one might argue that 
the fast transit times involved with short trajectories corresponds to large energy scales while the 
thermopower measures a local (at $E_F$) derivative. It would be interesting to investigate these effects 
theoretically. 

\begin{figure}[hb]
\begin{center}
\resizebox{5.5cm}{7cm}{\includegraphics{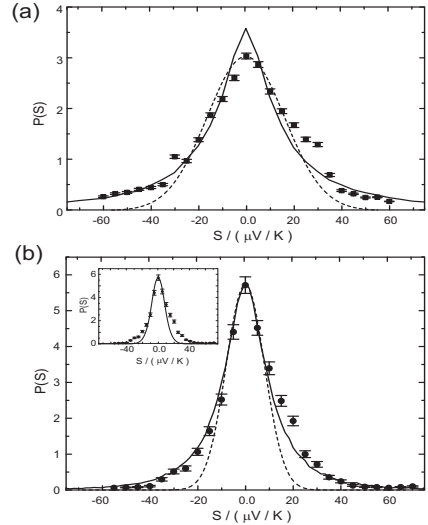}}
\end{center}
\begin{center}
\begin{minipage}{8cm}
\caption{(a) Thermopower distribution in presence of TRS ($|B|\leq 40$ mT). Experimental results (dots), 
simulation results (solid line) and Gaussian fit (dashed line) (b) Thermopower distribution for broken TRS 
($|B|\geq 50$ mT). Experimental results (dots), simulation results (solid line) and Gaussian fit (dashed 
line). Inset: Experimental thermopower distribution (dots) for broken TRS, compared to simulation results 
including strong dephasing (solid line).}\label{thermopowerdistribution}
\end{minipage}
\end{center}
\end{figure}

We now proceed to compare the statistics of the observed thermopower fluctuations with theoretical
predictions. The system symmetry, denoted in RMT with an integer $\beta$, changes with magnetic field; around 
$B=0$ T, time-reversal symmetry (TRS, $\beta=1$) is present while for 
higher magnetic fields this symmetry is broken ($\beta = 2$). The transition between the two regimes is a 
gradual one; for the present analysis we restrict ourselves to the two extremes. Experimentally, the regime 
with TRS is the magnetic field range where the weak localization effect dominates conductance measurements 
($|B| \leq 30 $ mT), TRS is broken for $|B|\geq 50$ mT. Counting the fluctuation amplitudes of the corrected 
thermopower for these two regimes leads to the histograms presented in Fig.~\ref{thermopowerdistribution}. 
The dashed lines are the best Gaussian fit to the experimental data. Clearly, strong deviations from Gaussian 
statistics occur.


These deviations are also expected from RMT calculation. In Ref.~\cite{Vanlangen} the thermopower 
distribution of a chaotic quantum dot has been obtained for single-mode contacts. The distribution exhibits a 
cusp at $S=0$ and tails as $P(S)\propto |S|^{-1-\beta} \ln |S|$, which displays a clear deviation from a 
Gaussian distribution. However, for a large number of conducting channels this distribution becomes again 
Gaussian. For the experimental data, taken with two spin-degenerate conducting modes in the leads, deviations 
from the a Gaussian distribution are still expected. Since in RMT an analytical treatment for the thermopower 
distribution is possible only for single-mode leads, Monte-Carlo simulations have to be employed for the 
present system. 
The Hamiltonian $\cal H$ of the closed dot is drawn randomly from the Gaussian ensemble

\begin{equation}
P({\cal H}) \propto {\rm e}^{-c{\rm Tr}\, {\cal HH}^\dagger}
\end{equation}

where $c$ is a constant setting the mean level spacing $\Delta$ and ${\cal H}$ is real symmetric ($\beta = 
1$) or hermitian ($\beta=2$). The scattering matrix \cite{Footnote} $M$ of the open system is calculated from 
$\cal H$ using \cite{Verbaarschot}

\begin{equation}
M(E) = 1 - 2\pi i W^\dagger\left(E-{\cal H}+i\pi WW^\dagger\right)^{-1} W.
\end{equation}

Here, $W$ is a rectangular matrix coupling the states in the dot to the scattering channels. The thermopower 
is then calculated from

\begin{equation}
\label{thermopower}
S=-\frac{\pi^2}{3}\frac{k_{\rm B}^2T}{e} \left. \frac{d}{dE} \ln T(E) \right|_{E=E_{\rm F}}
\end{equation}

where $T(E)=\sum_{\alpha\in 1, \beta\in 2} |M_{\alpha\beta}|^2$ is the probability for the transmission from 
lead $1$ to lead $2$. The differentiation is done numerically for each realization of the Hamiltonian. The 
density of states $\rho (E_F)= 1 / 2\pi i \; {\rm Tr}\, (M^\dagger dM/dE)|_{E=E_F}$ is used as a weight 
factor to account for a large charging energy \cite{Brouwer97b}. The resulting distributions of the 
thermopower fluctuations for both symmetry classes are shown as solid lines in 
Fig.~\ref{thermopowerdistribution}. Evidently, the simulations represent the experimental results much better 
than a Gaussian distribution function. 

For a comparision of experimental and theoretical distributions, the horizontal axes of both data sets have 
to be scaled, while keeping the normalization (area equals one). Only one scaling parameter was needed to fit 
the distributions for TRS and broken TRS (Fig.~3), which implies a temperature difference across the dot of 
$\Delta T = 56$ mK. As an independent check for the correctness of this scaling procedure, we have in 
addition measured the thermovoltage (for the same heating current) across QPC AD, adjusted for maximum 
thermovoltage, i.e. between the $N=1$ and $N=2$ conductance plateaus \cite{Molenkamp90}. For this 
configuration, the thermopower of a QPC is quantized, directly yielding a value for $\Delta T$. We find 
$\Delta T =54$ mK at $I=0.4$ $\mu$A, in good agreement with the scaling result.


It is possible to show that in thermopower measurements dephasing corrections, which are used to explain 
Gaussian conductance fluctuation distributions of chaotic quantum dots, are indeed irrelevant. The influence 
of a finite phase-coherence time $\tau_\phi$ can be studied by connecting the dot to a third virtual 
reservoir. The bath is coupled via $N_\phi\gg 1$ channels, corresponding to the incoherent limit 
($h/\tau_\phi\gg\Delta$). The chemical potential $\mu_\phi$ and temperature $T_\phi$ of the bath are chosen 
exactly in between those of the real reservoirs, such that it draws neither charge nor heat current. The 
distribution of the thermopower for broken TRS ($\beta=2$) then given by

\begin{eqnarray}
\label{incoherent}
P(\sigma ) &=& 2^{-8N^2+1} N {\rm e}^{-N|\sigma|}
\sum_{k=0}^{4N^2-1} \pmatrix{ 8N^2-2-k \cr 4N^2-1 } \frac{1}{k!} |2N\sigma|^k \nonumber \\ 
{\rm with}\: \sigma &=& \frac{3e}{\pi k_{\rm B}^2 T\Delta}
\left(\frac{\hbar}{\tau_\varphi} \right)^2 S. 
\end{eqnarray}

The result of Eq.~\ref{incoherent} is plotted for $N=2$ in the inset of Fig.~\ref{thermopowerdistribution}b 
(solid line). It can be seen that there is only little agreement between the experimental thermopower data 
and theory including dephasing which proofs that the observed statistics do not indicate the presence of any 
dephasing induced by electron heating. 


To conclude, we have demonstrated that the thermopower measurements on a chaotic quantum dot reveal directly 
the theoretically predicted non-Gaussian fluctuation distributions. In contrast to the best results obtained 
so far in conductance measurements, we don't have to include thermal broadening or dephasing to model our 
experimental results. Another distinction from conductance measurements is that the thermopower data are not 
influenced by weak localization or short trajectories. Therefore, thermopower measurements can be considered 
as an excellent tool in the area of investigating chaotic quantum transport properties in open systems. 

\acknowledgements
We thank C.W.J.\ Beenakker for his interest in this work. Part of this work was 
supported by the DFG, SFB-341 and DFG MO 771/3.

\end{multicols}


\begin{references}

\bibitem[*]{byline} Present and permanent address: Philips Research Laboratories, 5656 AA Eindhoven, The 
Netherlands.

\bibitem{Beenakker} C.W.J.\ Beenakker, Rev.\ Mod.\ Phys.\ {\bf69}, 731 (1997).

\bibitem{Marcus} C.M.\ Marcus, A.J.\ Rimberg, R.M.\ Westervelt, P.F.\ Hopkins, and A.C.\ Gossard, Phys.\ 
Rev.\ Lett.\ {\bf69}, 506 (1992).

\bibitem{Chan} I.H.\ Chan, R.M.\ Clarke, C.M.\ Markus, K.\ Campman, and A.C.\ Gossard, Phys.\ Rev.\ Lett.\ 
{\bf74}, 3876 (1995).

\bibitem{Baranger96} H.U.\ Baranger and P.A.\ Mello, Europhys.\ Lett.\ {\bf33}, 465 (1996).

\bibitem{PluharBaranger93} Z.\ Pluhar, H.A.\ Weidenm\"uller, J.A.\ Zuk, and C.H.\ Lewenkopf, Phys.\ Rev.\ 
Lett.\ {\bf73}, 2115 (1994); H.U.\ Baranger, R.A.\ Jalabert, and A.D.\ Stone, Phys.\ Rev.\ Lett.\ {\bf70}, 
3876 (1993).

\bibitem{Efetov95} K.B.\ Efetov, Phys.\ Rev.\ Lett.\ {\bf74}, 2299 (1995).

\bibitem{BBB} H.U.\ Baranger and P.A.\ Mello, Phys.\ Rev.\ B\ {\bf51}, 4703 (1995); P.W.\ Brouwer and C.W.J.\ 
Beenakker, Phys.\ Rev.\ B\ {\bf51}, 7739 (1995); P.W.\ Brouwer and C.W.J.\ Beenakker, Phys.\ Rev.\ B\ 
{\bf55}, 4695 (1997).

\bibitem{Huibers} A.G.\ Huibers, S.R.\ Patel, C.M.\ Marcus, P.W.\ Brouwers, C.I.\ Durn\"oz, and J.S.\ Harris, 
Jr.\ , Phys.\ Rev.\ Lett.\ {\bf 81}, 1917 (1998).

\bibitem{Molenkamp90} L.W.\ Molenkamp, H.\ van Houten, C.W.J.\ Beenakker, R.\ Eppenga, and C.T.\ Foxon, 
Phys.\ 
Rev.\ Lett.\ {\bf65}, 1052 (1990).

\bibitem{Molenkamp94} L.W.\ Molenkamp, A.A.M.\ Staring, B.W.\ Alphenaar, H.\ van Houten, and C.W.J.\ 
Beenakker, Semicond.\ Sci.\ Technol.\ {\bf 9}, 903 (1994); A.A.M.\ Staring, L.W.\ Molenkamp, B.W. Alphenaar, 
H.\ van Houten, O.J.A.\ Buyk, M.A.A.\ Mabesoone, C.W.J.\ Beenakker, and C.T.\ Foxon, Europhys.\ Lett.\ {\bf 
22}, 57 (1993).

\bibitem{Moeller} S.\ M\"oller, H.\ Buhmann, S.F.\ Godijn, and L.W.\ Molenkamp, Phys.\ Rev.\ 
Lett.\, issue of Nov.\ 30, 1998.

\bibitem{Fyodorov} Y.V.\ Fyodorov, Phys.\ Rev.\ Lett.\ {\bf 73}, 2688 (1994); Y.V. Fyodorov and A.D.\ Mirlin, 
Phys.\ Rev.\ B {\bf 51}, 13403 (1995).

\bibitem{Brouwer97b} P.W.\ Brouwer, S.A.\ van Langen, K.M.\ Frahm, M.\ B\"uttiker, and C.W.J.\ Beenakker, 
Phys.\ Rev.\ Lett.\ {\bf 79}, 913 (1997).

\bibitem{Vanlangen} S.A.\ van Langen, P.G.\ Silvestrov, and C.W.J.\ Beenakker, Superlatt.\ and Microstr.\ 
{\bf 23}, 691 (1998).

\bibitem{average} Small fluctuation are still visible both in the averaged conductance (Fig.~1b) and the 
averaged thermovoltage (inset of Fig.~2b) due to the limited range of gate voltage $V_g^B$ that is accessible 
in the experiment without destroying the sample. 

\bibitem{kearney} M.J.\ Kearney, R.T.\ Stone, and M.\ Pepper, Phys.\ Rev.\ Lett.\ {\bf 66}, 1622 (1991).

\bibitem{Footnote} In literature the symbol S is normally used for the scattering matrix. In this paper we 
use the symbol M to avoid confusion with the thermopower S.

\bibitem{Verbaarschot} J. J. M. Verbaarschot, H. A. Weidenmuller, and M. R. Zirnbauer, Phys.\ Rep.\ {\bf 
129}, 367 (1985).

\end{references}
\end{document}